\documentclass[aps,prl,reprint,superscriptaddress]{revtex4-2}

\usepackage{graphicx}
\usepackage{amsmath}
\usepackage{amssymb}
\usepackage{bm}
\usepackage{xcolor}
\usepackage[colorlinks=true, breaklinks=true, linkcolor=blue, citecolor=blue, urlcolor=blue]{hyperref}

\newcommand{\Heff}{H_{\mathrm{eff}}}
\newcommand{\Unonadd}{U^{(\mathrm{nonadd})}_{bxA}}
\newcommand{\Upol}{U^{(\mathrm{pol})}_{bxA}}
\newcommand{\Pspace}{\mathcal{P}}
\newcommand{\Qspace}{\mathcal{Q}}
\newcommand{\PA}{P_A}
\newcommand{\QA}{Q_A}
\newcommand{\Pbx}{P_{bx}}
\newcommand{\Qbx}{Q_{bx}}
\newcommand{\bra}[1]{\langle #1 |}
\newcommand{\ket}[1]{| #1 \rangle}

\begin{document}

\title{Dynamical Origin of Spectroscopic Quenching in Knockout Reactions}

\author{Jin Lei}
\email[]{jinl@tongji.edu.cn}
\affiliation{School of Physics Science and Engineering, Tongji University, Shanghai 200092, China.}

\date{\today}

\begin{abstract}
Nucleon-removal reactions are a primary tool for extracting
single-particle structure of rare isotopes, yet the ratio
$R_s=\sigma_{\exp}/\sigma_{\mathrm{th}}$ of measured to theoretical
cross sections drops systematically below unity for deeply bound
nucleons. I derive the exact effective three-body Hamiltonian for
composite-projectile reactions using a sequential double Feshbach
projection and show that the standard additive model misses two induced
interactions: a non-additive term from virtual target excitations and a
polarization potential from excluded projectile configurations. Their
omission overestimates the stripping cross section, producing apparent
quenching distinct from genuine nuclear-structure correlations. This
mechanism offers a dynamical origin for the strong separation-energy
dependence of the quenching ratio, a feature unique to knockout analyses. Existing four-body CDCC
calculations for $^{6}$Li validate the framework: the proper Feshbach
reference reproduces elastic scattering data, while a phenomenological
optical potential double counts the breakup absorption and fails.
\end{abstract}

\maketitle

\textit{Introduction.}---Nucleon-removal reactions at intermediate
energies have transformed nuclear spectroscopy, providing direct access
to single-particle structure from light $p$-shell systems to the most
neutron-rich isotopes produced at modern rare-isotope
facilities~\cite{Navin1998,Bazin2003,Steppenbeck2013,Chen2019,Taniuchi2019,Frotscher2020,Otsuka2020}.
Yet a long-standing puzzle undermines the reliability of the
spectroscopic information so extracted: the ratio
$R_s=\sigma_{\exp}/\sigma_{\mathrm{th}}$ of measured to theoretical
cross sections drops systematically from near unity for loosely bound
nucleons to ${\sim}0.25$ for deeply bound ones, with a pronounced
dependence on the separation-energy asymmetry
$\Delta S$~\cite{Gade2008,Lee2010,Jensen2011,Tostevin2014,Tostevin2021,Aumann2021}.
If structural in origin, this quenching would imply a dramatic
breakdown of the independent-particle picture for nuclei with large
proton--neutron imbalance, a conclusion with far-reaching consequences
for the nuclear-structure program at FRIB, RIBF, and GSI.

The strong $\Delta S$ dependence, however, is unique to eikonal
knockout analyses~\cite{Tostevin2014,Tostevin2021}. Transfer reactions
yield $R_s\approx 0.5$--$0.7$ with essentially no asymmetry dependence~\cite{Flavigny2013}; quasifree $(p,2p)$
scattering shows isospin-independent reduction
factors~\cite{Atar2018}; $(e,e'p)$ data exhibit a uniform
${\sim}0.6$--$0.7$ depletion~\cite{Kramer2001,Kay2013}. That different probes
applied to the same nuclei yield different spectroscopic factors is a
clear signature that the quenching encodes, at least in part, a
deficiency in the reaction description rather than a property of the
nuclear wave function.

A key observation is that independent eikonal,
quantum-mechanical, and transfer-to-continuum
implementations of the standard model yield mutually
consistent $\sigma_{sp}$~\cite{LeiBonaccorso2021,GomezRamos2021},
pointing to a deficiency in the model Hamiltonian itself rather than in
the dynamical approximation. The
standard factorization lacks a controlled derivation from the
underlying many-body problem: two-body optical potentials are constructed in a
two-body context, and their validity as effective interactions inside a
three-body Hamiltonian cannot be taken for granted.
The problem is most acute for deeply bound nucleon
removal, where the reaction probes the nuclear interior;
nonsudden effects~\cite{Flavigny2012}, core-excitation
pathways~\cite{Moro2012}, core-destruction
processes~\cite{Bertulani2023}, and continuum-coupling
effects~\cite{Xie2026} have each been shown to modify $\sigma_{sp}$ in
this regime. Any overestimate of
$\sigma_{sp}$ inflates $\sigma_{\mathrm{th}}$ and drives $R_s$ below
unity, mimicking genuine quenching. The resulting strong
$\Delta S$ dependence is, at least in part, a dynamical artifact distinct
from the uniform depletion observed in $(e,e'p)$.

Formal justification for the effective few-body description has a long
history~\cite{Timofeyuk2020}. For the deuteron case, projecting onto a
three-body model generates spectator-dependent energy shifts and a
genuine three-body
interaction~\cite{Austern1968,KozackLevin1986,Polyzou1979}. Subsequent
work quantified these
corrections~\cite{Timofeyuk2013PRL,Johnson2014,Timofeyuk2013PRC,Bailey2017}
and showed that they modify transfer cross sections appreciably.
However, these analyses were restricted to $A(d,p)$ and the explicit
connection to spectroscopic quenching was not established.

In this Letter I generalize these earlier results to arbitrary
composite projectiles and establish the direct link between the missing
induced interactions and the observed spectroscopic quenching. Using a
double Feshbach projection I derive the exact effective three-body
Hamiltonian, whose decomposition identifies every term that the
standard additive model neglects and provides a practical
self-consistency criterion for distinguishing genuine structure
depletion from dynamical artifacts.

\begin{figure}[t]
\centering
\includegraphics[width=\columnwidth]{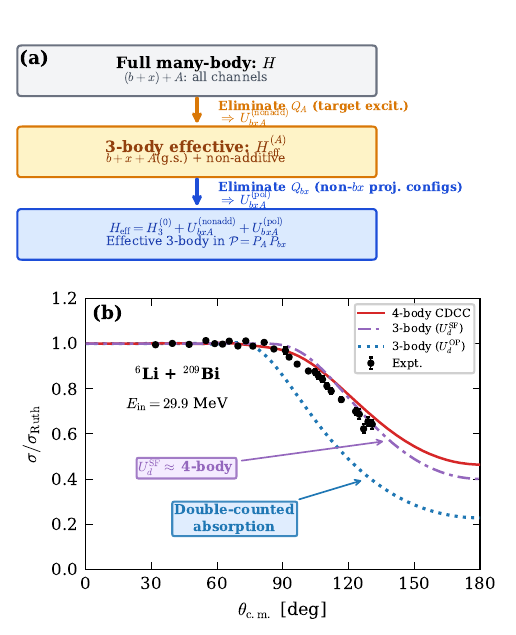}
\caption{(a)~Sequential double Feshbach projection: eliminating target
excitations ($\QA$) yields $\Unonadd$; eliminating excluded projectile
configurations ($\Qbx$) yields $\Upol$.
(b)~$^{6}$Li~$+$~$^{209}$Bi elastic scattering at
$E_{\mathrm{in}}=29.9$~MeV. Data from
Ref.~\cite{Keeley2003}; theory from
Ref.~\cite{Watanabe2012}. Four-body CDCC (solid red) and three-body
CDCC with $U_d^{\mathrm{SF}}$ (dot-dashed purple) reproduce the data.
Three-body CDCC with $U_d^{\mathrm{OP}}$ (dotted blue) underestimates
the elastic cross section: the $d$-breakup DPP is double counted,
producing excess absorption.}
\label{fig:channels}
\end{figure}

\textit{Double Feshbach projection.}---I consider a projectile $a$ resolved
into clusters $b$ and $x$ scattering from a target $A$, with
Jacobi coordinates $\mathbf{r}$ for $b$--$x$ relative motion and
$\mathbf{R}$ for $a$--$A$ relative motion.
Antisymmetrization between projectile and target nucleons
is subsumed into the effective interactions, as is standard in
few-body reaction theory~\cite{Austern1970}.
The full Hamiltonian is $H = T_R + T_r + H_A + H_a
+ V_{bA} + V_{xA}$, where $H_A$ is the target internal Hamiltonian,
$H_a = H_b + H_x + V_{bx}$ collects the projectile internal
interactions (with $H_b$, $H_x$ the fragment internal Hamiltonians
and $V_{bx}$ the inter-constituent interaction that binds the
projectile; the projectile eigenstates satisfy
$(T_r{+}H_a)|\phi_{bx}^{(\nu)}\rangle=\varepsilon_\nu|\phi_{bx}^{(\nu)}\rangle$), and
$V_{bA}$, $V_{xA}$ contain all pairwise interactions between the
constituents of each fragment and the target. I project onto the three-body model space using two
commuting projectors that act on independent subspaces:
\begin{equation}\label{eq:projectors}
  \Pspace = \PA\,\Pbx, \qquad \Qspace = 1-\Pspace,
\end{equation}
where $\PA = \ket{\phi_A}\bra{\phi_A}$ projects onto the target ground state
and the projectile-space projector is defined as
$\Pbx=P_b^{\mathrm{int}}P_x^{\mathrm{int}}P_{bx}^{\mathrm{rel}}$, where
$P_b^{\mathrm{int}}$ and $P_x^{\mathrm{int}}$ retain the chosen internal
states of fragments $b$ and $x$, and $P_{bx}^{\mathrm{rel}}$ retains the
selected $b{+}x$ relative-motion channels (bound and continuum,
or discretized continuum as in CDCC). Its complement $\Qbx$
encompasses all configurations outside
this space, including those in which the projectile does not have the
$b{+}x$ structure. Eliminating $\Qspace$-space exactly yields the
Feshbach effective
Hamiltonian~\cite{Feshbach1958,Feshbach1962,Mahaux1969}
\begin{equation}\label{eq:Heff}
  \Heff(E) = \Pspace H\Pspace
  + \Pspace H\Qspace\,\frac{1}{E{+}i\epsilon - \Qspace H\Qspace}\,
  \Qspace H\Pspace.
\end{equation}
Here $\Heff$ is exact, energy-dependent,
nonlocal, and non-Hermitian; its anti-Hermitian part accounts for flux
loss into channels excluded from~$\Pspace$.

\textit{Decomposition of the effective interaction.}---To expose the
multi-body content, I split each fragment--target interaction into a
reference optical potential and a residual coupling,
$V_{bA}=U_{bA}+\Delta v_{bA}$ and similarly for $xA$. Here $U_{bA}$
is not a specific phenomenological or microscopic potential but an
as-yet-unspecified reference: any choice defines a valid decomposition,
and $\Delta v$ absorbs whatever $U$ does not capture. The
decomposition is exact for every $U$; the choice only affects the
relative size of the induced terms.
A particularly useful choice is one
satisfying the projected Feshbach condition
$P_b^{\mathrm{int}}\PA U_{bA}\PA P_b^{\mathrm{int}}
=P_b^{\mathrm{int}}\PA V_{bA}\PA P_b^{\mathrm{int}}$
(and similarly with $P_x^{\mathrm{int}}$ for $xA$),
which projects onto both the target and fragment ground states. For
structureless fragments (e.g., a nucleon), $P_b^{\mathrm{int}}$ reduces
to unity and this is the standard Feshbach condition; for composite
fragments it yields a single-folding potential
(see End Matter). This condition eliminates
the first-order non-additive correction in the model space. Performing the Feshbach
elimination sequentially, first eliminating target excitations
($\QA$-space), then projecting onto the projectile model space
($\Pbx$), yields the exact two-term decomposition (see End Matter for detailed derivation)
\begin{equation}\label{eq:decomposition}
  \Heff(E) = H_3^{(0)}
  + \Unonadd(E)
  + \Upol(E),
\end{equation}
where $H_3^{(0)}=T_R+T_r+H_a+U_{bA}+U_{xA}$ is the
conventional additive three-body Hamiltonian, with $U_{bA}$ and $U_{xA}$ the reference potentials defined above. Most reaction calculations
use $H_3^{(0)}$ alone with the standard factorization
\begin{equation}\label{eq:factorized}
  \sigma_{\mathrm{th}} = \sum_\beta S_\beta\,\sigma_{sp}^{(\beta)}[U_{bA},U_{xA}],
\end{equation}
thereby neglecting both induced terms.

The first induced term, $\Unonadd$
[End Matter, Eq.~(\ref{eq:Unonadd_full})], is the non-additive
effective interaction generated by
eliminating target excitations ($\QA$-space), the generalization to
composite projectiles of the three-body effective interaction identified
for the deuteron case
in~\cite{Austern1968,Polyzou1979,KozackLevin1986}. With the
Feshbach-condition choice of $U$, the model-space projection
$\Pbx\bra{\phi_A}\Delta V\ket{\phi_A}\Pbx=0$
($\Delta V=\Delta v_{bA}+\Delta v_{xA}$); the off-diagonal remainder
$\Pbx\bra{\phi_A}\Delta V\ket{\phi_A}\Qbx$ is absorbed into $\Upol$
through the second projection step.
The irreducible three-body
content of $\Unonadd$ resides in the cross terms
\begin{align}\label{eq:cross}
  &\bra{\phi_A}\Delta v_{bA}\,\QA G_A\,\Delta v_{xA}\ket{\phi_A} \nonumber\\
  &\quad + \bra{\phi_A}\Delta v_{xA}\,\QA G_A\,\Delta v_{bA}\ket{\phi_A},
\end{align}
with $G_A=(E{+}i\epsilon - \QA H\QA)^{-1}$. These cross terms
couple the two fragments through virtual target excitations and
constitute the microscopic origin of non-additivity. Even if
$U_{bA}$ and $U_{xA}$ are individually exact, their sum does not
reproduce the correct interaction once intermediate target
excitations are eliminated~\cite{Austern1968,Polyzou1979}. For $a=d$ this reduces
exactly to the induced three-body potential $U_{npA}$ of
Refs.~\cite{Timofeyuk2013PRL,Timofeyuk2013PRC,Johnson2014}; the
operator-energy dependence [Eq.~(\ref{eq:operator_energy})] yields the
adiabatic energy-shift prescription of Ref.~\cite{Johnson1970} as its
leading approximation.

The second induced term, $\Upol$
[End Matter, Eq.~(\ref{eq:Upol_full})], is the dynamical
polarization potential~\cite{Feshbach1962} arising from eliminating
projectile configurations outside the model space ($\Qbx$-space) within
the target-projected effective Hamiltonian
$\Heff^{(A)}=H_3^{(0)}+\Unonadd$. Because the target excitations have
already been eliminated, $\Upol$ avoids double counting with $\Unonadd$.
The $\Qbx$-space comprises both $b{+}x$ continuum states
beyond those retained in $\Pbx$ and non-$b{+}x$ configurations; when the model
space already includes the $b{+}x$ continuum (as in CDCC), the residual
$\Upol$ is dominated by the non-$b{+}x$ component and encodes the
physics traditionally compressed into a spectroscopic factor.
Together, $\Unonadd$ and $\Upol$
represent everything that the standard additive model $H_3^{(0)}$ misses.

\textit{Spectroscopic quenching from missing interactions.}---The standard
practice [Eq.~(\ref{eq:factorized})] uses $H_3^{(0)}$ alone, neglecting both
$\Unonadd$ and $\Upol$. These missing terms carry imaginary parts that
provide additional absorption in the model space; when the dominant
omitted absorption acts in the residue-survival channel, the additive
model overestimates $\sigma_{sp}$
[End Matter, Eq.~(\ref{eq:delta_sigma_neg})], producing $R_s<1$ even
when the underlying spectroscopic factor is correct.
This deficiency is structural: the cross
terms~(\ref{eq:cross}) are irreducibly three-body and cannot be
generated by any additive combination of two-body optical
potentials~\cite{KoningDelaroche2003,JLM1977}, no matter how
accurately each is constrained.

The effect is most pronounced for deeply bound nucleon removal, where
the reaction probes shorter distances and $\Unonadd$, $\Upol$ are
largest. The probe dependence of $R_s$
(knockout shows strong $\Delta S$
dependence~\cite{Tostevin2014,Tostevin2021}; transfer is much
flatter~\cite{Flavigny2013}) follows naturally: each framework retains a
different $\Pspace$-space and misses a different portion of $\Unonadd$
and $\Upol$. The eikonal knockout model, lacking channel coupling,
omits both terms entirely.

In the present formalism, ``quenching'' ($R_s<1$) does not necessarily
imply that the structure-model spectroscopic factor $S_\beta$ is wrong;
it can equally reflect an overestimated $\sigma_{sp}$ from the missing
induced terms. The decomposition~(\ref{eq:decomposition}) provides a
practical self-consistency criterion. In Route~A, the reaction dynamics
is computed with $\Heff$ (or its dynamical equivalents, i.e., channel
coupling that generates the induced interactions explicitly); the
spectroscopic factor enters only as the overlap norm
$S_P=\int d^3r\,|I_{bx}(\mathbf{r})|^2$
[End Matter, Eq.~(\ref{eq:SP_def})], and no additional reduction
should be applied, since the flux loss into excluded configurations is
already accounted for by $\Upol$. In Route~B, the dynamics is
approximated by $H_3^{(0)}$ alone; the missing absorption from
$\Unonadd$ and $\Upol$ is then implicitly absorbed into an effective
spectroscopic factor $S_{\mathrm{eff}}<S_P$ that compensates the
overestimated $\sigma_{sp}$. This effective SF is explicitly model
dependent and should not be compared directly to a structure-model
overlap norm (see End Matter for formal discussion).

\textit{$^{6}$Li benchmark.}---For nucleon knockout
($a=\mathrm{core}+N$), evaluating the induced interactions would require a
coupled many-body treatment of virtual target excitations and all
excluded projectile configurations (core-excited and non-cluster states),
a problem intractable with present methods.
$^{6}$Li provides the ideal surrogate: the tight binding of the
$\alpha$ particle (${\sim}28$~MeV) makes the $\alpha+d$ clustering
a controlled approximation, so $^{6}\mathrm{Li}\approx\alpha+d$ with
$d=n+p$. The transition from a four-body ($\alpha+n+p+A$) to a
three-body ($\alpha+d+A$) description then generates precisely the
induced interactions identified above. Published four-body CDCC
calculations~\cite{Watanabe2012,Watanabe2015} provide a quantitative test
without requiring new computations. In the $\alpha+d+A$
model space, the effective Hamiltonian takes the form
$\Heff^{(\alpha d)}=T_R+T_r+H_{^6\mathrm{Li}}+U_{\alpha A}+U_{dA}
+U^{(A)}_{\alpha dA}+U^{(^6\mathrm{Li})}_{\alpha dA}$, where the two
induced terms correspond directly to the general
decomposition~(\ref{eq:decomposition}): $U^{(A)}_{\alpha dA}$ is
$\Unonadd$ for this system, the non-additive interaction generated by
eliminating target excitations, while
$U^{(^6\mathrm{Li})}_{\alpha dA}$ is $\Upol$, the polarization
potential from eliminating non-$\alpha{+}d$ configurations of $^{6}$Li.

The projected Feshbach condition~(\ref{eq:proj_Feshbach})
applied to $\xi=d$ gives
$\bra{\phi_d}\bra{\phi_A}V_{dA}\ket{\phi_A}\ket{\phi_d}=U_{dA}$,
i.e., the first-order (Hartree) folding of $V_{dA}$, and eliminates
the $dA$ contribution to the first-order mismatch, leaving only the
second-order contributions
(diagonal DPP renormalization and irreducible cross terms). The
standard single-folding potential
$U_d^{\mathrm{SF}}=\langle\phi_d|U_{nA}{+}U_{pA}|\phi_d\rangle$
goes further. Since $U_{nA}$ and $U_{pA}$ are themselves Feshbach
optical potentials (evaluated at half the deuteron incident
energy~\cite{Johnson1970}),
$U_d^{\mathrm{SF}}$ additionally absorbs the diagonal DPP
renormalization into the reference. The residual $\Unonadd$ is then
dominated by the irreducible cross terms~(\ref{eq:cross}) and
operator-energy
corrections~\cite{Timofeyuk2013PRC,Bailey2017}. By contrast, the phenomenological deuteron optical potential
$U_d^{\mathrm{OP}}$, fitted to free $d{+}A$ elastic data,
absorbs the $d$-breakup DPP
into the reference. This absorption will be regenerated
dynamically by the CDCC coupling, and $U_d^{\mathrm{OP}}$ therefore violates the Feshbach condition. Since
$U_{\alpha A}$ approximately satisfies the condition for the tightly
bound $\alpha$,
$\bra{\phi_A}\Delta v_{\alpha A}\ket{\phi_A}\approx 0$ and the
dominant first-order mismatch resides in the $dA$
channel alone. With $\Delta v_{dA}=V_{dA}-U_d^{\mathrm{OP}}$,
\begin{align}\label{eq:correction}
  \bra{\phi_d}\bra{\phi_A}\Delta v_{dA}\ket{\phi_A}\ket{\phi_d}
  &= \bra{\phi_d}\bra{\phi_A}V_{dA}\ket{\phi_A}\ket{\phi_d}
  - U_d^{\mathrm{OP}} \nonumber\\
  &\approx U_d^{\mathrm{SF}} - U_d^{\mathrm{OP}}
  = -\Delta U_{\mathrm{DPP}}^{(np)},
\end{align}
where $\bra{\phi_d}\bra{\phi_A}V_{dA}\ket{\phi_A}\ket{\phi_d}\approx U_d^{\mathrm{SF}}$
(see End Matter); the $d$-breakup DPP
$\Delta U_{\mathrm{DPP}}^{(np)}$ is the dominant uncompensated
mismatch, a substantial quantity for the weakly bound deuteron
($\varepsilon_d=2.22$~MeV).

Watanabe \textit{et al.}~\cite{Watanabe2012} tested this quantitatively
for $^{6}$Li$+{}^{209}$Bi elastic scattering at $E_{\mathrm{in}}=29.9$
and $32.8$~MeV; Fig.~\ref{fig:channels}(b) shows the $29.9$~MeV case. Four-body CDCC
($n{+}p{+}\alpha{+}{}^{209}$Bi) serves as benchmark. Three-body CDCC with
$U_d^{\mathrm{SF}}$, which approximates the proper Feshbach
reference (see discussion above), reproduces the
four-body result with no adjustable parameters, while three-body CDCC with
$U_d^{\mathrm{OP}}$ strongly underestimates the elastic cross section
and fails to reproduce the data: the $d$-breakup DPP already embedded
in $U_d^{\mathrm{OP}}$ is double counted by the CDCC $\alpha$--$d$
coupling, producing excess absorption that removes too much flux from
the elastic channel. This is Eq.~(\ref{eq:decomposition}) at work: $U_d^{\mathrm{SF}}$
realizes Route~A (proper Feshbach reference, induced interactions
generated dynamically by the CDCC coupling), while $U_d^{\mathrm{OP}}$
violates the condition and double counts the breakup absorption.

The comparison illustrates two complementary manifestations of the
same root cause, inconsistent treatment of the induced terms in
Eq.~(\ref{eq:decomposition}). CDCC with $U_d^{\mathrm{OP}}$
(overcomplete reference) double counts the $d$-breakup absorption and
removes too much elastic flux [dotted blue in
Fig.~\ref{fig:channels}(b)]; the standard additive model $H_3^{(0)}$
(no channel coupling) omits both induced terms, underestimates
absorption, and overestimates $\sigma_{sp}$
[End Matter, Eq.~(\ref{eq:delta_sigma_neg})]. The $^{6}$Li data
directly confirm the first failure; the second is the prediction
of the present formalism for knockout. The proper Feshbach reference
$U_d^{\mathrm{SF}}$ (Route~A) avoids both.

A further insight from the four-body analysis~\cite{Watanabe2015} is
$d\alpha$ dominance: selectively disabling non-$d{+}\alpha$
configurations in the four-body CDCC calculation has negligible effect
on the elastic cross section, confirming that $^{6}$Li breakup is
overwhelmingly into $d{+}\alpha$ rather than $n{+}p{+}\alpha$. In the
language of the present formalism this means $\Upol\approx 0$: even
though the $d{+}\alpha$ spectroscopic factor
$S_{d\alpha}\approx 0.70$~\cite{Kikuchi2011} leaves ${\sim}30\%$ of
the wave function in non-$d{+}\alpha$ configurations ($\Qbx$-space),
their dynamical coupling to the elastic channel is weak, and
the missing correction is dominated by $\Unonadd$ alone. The leading
mismatch between $U_d^{\mathrm{OP}}$ and $U_d^{\mathrm{SF}}$ is the
$d$-breakup DPP [Eq.~(\ref{eq:correction})]; the remaining irreducible
cross terms~(\ref{eq:cross}) are small, as confirmed by the close
agreement of three-body CDCC with $U_d^{\mathrm{SF}}$ and the
four-body benchmark, and as expected from the tight binding of the
$\alpha$ particle (${\sim}28$~MeV), which keeps $\Delta v_{\alpha A}$
small.

\textit{Implications.}---The general
decomposition~(\ref{eq:decomposition}) shows that the standard additive
model $H_3^{(0)}$ is systematically incomplete: for any composite projectile
$a=b+x$, it misses both $\Unonadd$ and $\Upol$, whose magnitudes are
system-dependent but generically nonzero. Because these missing terms
modify $\sigma_{sp}$, the observed spectroscopic quenching ($R_s<1$) need
not originate solely from nuclear-structure correlations. Instead, the
formalism identifies three distinct contributions to $R_s<1$: (i) genuine
structure depletion (the overlap-function norm reduced by correlations),
(ii) overestimated $\sigma_{sp}$ from neglected induced interactions, and
(iii) model inconsistency when phenomenological optical potentials already
absorb part of the missing physics. For deeply bound nucleon removal, where
contributions (ii) and (iii) are largest, the apparent quenching can be
substantially influenced, and in some systems dominated, by dynamical incompleteness rather than by the spectroscopic
factor itself. The probe dependence of $R_s$ follows directly, as
discussed above: each framework overestimates $\sigma_{sp}$ by a
different amount depending on which induced terms it omits.
Quoted SFs should always be tied to the
reaction model space and interaction set with which they were extracted.

Two caveats delimit the scope. First, the present mechanism
is specific to reactions involving composite projectiles;
electromagnetic probes such as
$(e,e'p)$~\cite{Kramer2001,Kay2013} involve no composite projectile
and lie outside the present framework.
For hadronic knockout, the induced interactions are
largest where the reaction probes short distances, i.e., for deeply
bound nucleon removal, while they become negligible for loosely bound
orbits near the nuclear surface, naturally producing the strong
$\Delta S$ dependence unique to eikonal knockout analyses.
Second, computing the induced interactions for
nucleon knockout remains intractable; the $^{6}$Li test validates the
mechanism in the only system where a controlled
comparison is presently feasible.
Quantitative calibration remains an essential next step.

The framework is prescriptive. For nucleon knockout from a composite
projectile ($a=\mathrm{core}{+}N$), the cross terms~(\ref{eq:cross})
couple the removed nucleon and the core through virtual excitations of
the target; these are absent from any additive combination of
individual optical potentials $U_{NA}+U_{\mathrm{core},A}$, no matter how
accurately each is determined. If coupled-channel core-excitation
coupling is employed to generate part of the missing dynamics, the
reference optical potential for the core--target system must exclude the
core-excitation DPP to avoid double counting, precisely the lesson of
the $U_d^{\mathrm{SF}}$ versus $U_d^{\mathrm{OP}}$ comparison.
The same logic extends to transfer, $(p,pN)$, and other clusterized
projectiles ($^{7}$Li, $^{11}$Be, \ldots), though the relative
importance of $\Unonadd$ vs.\ $\Upol$ will vary with the binding
energies and structure of the constituents. In particular,
the $^{6}$Li benchmark is dominated by $\Unonadd$ ($d\alpha$
dominance makes $\Upol\approx 0$), but for standard nucleon knockout
($a=\mathrm{core}+N$) the hierarchy may well be inverted: the
$\Qbx$-space, comprising all core-excited and non-cluster
configurations, is vast, and $\Upol$ is expected to be the
leading missing term (see End Matter for a detailed discussion).
Since both terms are irreducibly three-body, they cannot be
represented as corrections to any single two-body potential.
Extending such controlled evaluations to nucleon knockout
systems~\cite{Watanabe2012,Watanabe2015} is the natural next step
toward placing reduction-factor extractions on a
self-consistent footing.

\begin{acknowledgments}
The author thanks S.~Watanabe for providing the four-body CDCC
calculation results. This work was supported by the National Natural
Science Foundation of China (Grant Nos.~12475132 and 12535009) and the
Fundamental Research Funds for the Central Universities.
\end{acknowledgments}

\bibliography{references}


\appendix*
\onecolumngrid
\section*{End Matter}

\setcounter{equation}{0}
\renewcommand{\theequation}{A\arabic{equation}}

\textit{Detailed derivation of the effective three-body Hamiltonian.}---I
provide a self-contained derivation of the decomposition stated in
Eq.~(\ref{eq:decomposition}) of the main text. Here
$\Pbx=P_b^{\mathrm{int}}P_x^{\mathrm{int}}P_{bx}^{\mathrm{rel}}$ is the
projectile-space projector as defined in Eq.~(\ref{eq:projectors}).

The $\Qspace$-space complement of $\Pspace=\PA\Pbx$ decomposes as
$\Qspace = \QA\Pbx + \PA\Qbx + \QA\Qbx$, reflecting three classes of
excluded configurations: target excitations with the projectile retained
($\QA\Pbx$), excluded projectile configurations with the target in its
ground state ($\PA\Qbx$), and configurations where both target and
projectile are excited ($\QA\Qbx$). The $\Pspace H\Qspace$ coupling
therefore has three independent channels, and the resolvent
$(E+i\epsilon-\Qspace H\Qspace)^{-1}$ couples them.

I write each fragment--target interaction as
$V_{bA}=U_{bA}+\Delta v_{bA}$ and $V_{xA}=U_{xA}+\Delta v_{xA}$, where
$U_{bA}$ and $U_{xA}$ are reference optical potentials. The
decomposition is exact for any choice of $U$. To define the
Feshbach condition, introduce fragment-internal projectors
$P_b^{\mathrm{int}}=\ket{\phi_b}\bra{\phi_b}$ and
$P_x^{\mathrm{int}}=\ket{\phi_x}\bra{\phi_x}$ (for whichever fragment is
composite; unity for structureless fragments).
When $b$ is composite, $\PA V_{bA}\PA$ remains an operator in
$b$'s internal coordinates; the projected Feshbach condition
\begin{equation}\label{eq:proj_Feshbach}
  P_\xi^{\mathrm{int}}\,\PA U_{\xi A}\PA\,P_\xi^{\mathrm{int}}
  =P_\xi^{\mathrm{int}}\,\PA V_{\xi A}\PA\,P_\xi^{\mathrm{int}},
  \qquad \xi=b,x,
\end{equation}
projects onto both the target and fragment ground states, yielding
single-folding potentials
$U_\xi^{\mathrm{SF}}\approx
P_\xi^{\mathrm{int}}\,\PA V_{\xi A}\PA\,P_\xi^{\mathrm{int}}$
as the practical reference.
The first-order mismatch for a general reference
$U_{\xi A}^{\mathrm{ref}}$ is then
\begin{equation}
  P_\xi^{\mathrm{int}}\,\PA\Delta v_{\xi A}\PA\,P_\xi^{\mathrm{int}}
  =U_\xi^{\mathrm{SF}}-U_{\xi A}^{\mathrm{ref}},
\end{equation}
which vanishes for the single-folding choice and is nonzero for a
phenomenological optical potential that absorbs additional dynamics
(e.g., breakup DPP).
For the $^{6}$Li benchmark ($a=\alpha+d$), the relevant nontrivial channel is
$\xi=d$, yielding the correction in Eq.~(\ref{eq:correction}); the $\alpha$
channel mismatch is small.

\textit{Real vs.\ complex reference potentials.}---Because $V_{bA}$ is
Hermitian, the projected Feshbach
condition~(\ref{eq:proj_Feshbach}) is naturally satisfied by a real
reference potential; all absorption (flux loss into excluded channels) is then
generated dynamically by the imaginary parts of $\Unonadd$ and $\Upol$.
If instead one adopts a complex optical potential as reference, as is
common in practice, its imaginary part already accounts for some
absorption, and $\Delta v$ adjusts accordingly. Since the bare interaction
$V$ is real, $\Delta v = V - U$ acquires the opposite imaginary part,
$\mathrm{Im}(\Delta v) = -\mathrm{Im}(U)$, which partially cancels the
absorption built into $U$ and redistributes it into the induced terms.
The total $\Heff$ is independent of this choice, but the bookkeeping
differs: with a real $U$, $H_3^{(0)}$ is Hermitian and all non-Hermiticity
resides in the induced terms; with a complex $U$, $H_3^{(0)}$ is already
non-Hermitian and the induced terms provide corrections. In the $^{6}$Li
benchmark, $U_d^{\mathrm{SF}}$ is computed from complex nucleon optical
potentials $U_{nA}$ and $U_{pA}$ (themselves Feshbach optical potentials of
the nucleon--target subsystems), so $U_d^{\mathrm{SF}}$ is complex. The
phenomenological $U_d^{\mathrm{OP}}$, however, absorbs additional
$d$-breakup DPP into its imaginary part, violating the Feshbach condition
and generating a nonvanishing first-order correction~(\ref{eq:correction}).

Because the projected Feshbach condition ensures
$\Pbx\PA\Delta v_{bA}\PA\Pbx=\Pbx\PA\Delta v_{xA}\PA\Pbx=0$, the
$\Pspace H\Pspace$ block reduces to
\begin{equation}\label{eq:H30}
  \Pspace H\Pspace = \Pbx\,H_3^{(0)}\,\Pbx,
\end{equation}
i.e., $H_3^{(0)}$ restricted to the model space $\Pspace$.

The $\Pspace H\Qspace$ coupling separates into contributions from target
excitations and from projectile-space truncation. The target-excitation
channel gives $\Pspace H\,\QA\Pbx = \Pbx\PA(\Delta v_{bA}+\Delta
v_{xA})\QA\Pbx$, which, after projection through the $\QA$-space
resolvent, yields the non-additive interaction:
\begin{align}\label{eq:Unonadd_full}
  \Unonadd(E) &= \bra{\phi_A}\Delta V\ket{\phi_A}
  \nonumber\\
  &\quad + \bra{\phi_A}\Delta V\,\QA
  \frac{1}{E{+}i\epsilon - \QA H\QA}\QA\,\Delta V\ket{\phi_A},
\end{align}
where $\Delta V=\Delta v_{bA}+\Delta v_{xA}$. Expanding the quadratic
form in $\Delta V$ produces four terms: the diagonal contributions
$\bra{\phi_A}\Delta v_{bA}\QA G_A\Delta v_{bA}\ket{\phi_A}$ and
$\bra{\phi_A}\Delta v_{xA}\QA G_A\Delta v_{xA}\ket{\phi_A}$ renormalize
the individual optical potentials $U_{bA}$ and $U_{xA}$ into the full
(energy-dependent, nonlocal) optical interactions
$\mathcal{U}_{bA}$ and $\mathcal{U}_{xA}$, while the cross terms~(\ref{eq:cross})
are irreducibly three-body and constitute the non-additive content of
$\Unonadd$.
With the projected Feshbach condition~(\ref{eq:proj_Feshbach}),
the first term in Eq.~(\ref{eq:Unonadd_full}) vanishes in the model
space, $\Pbx\bra{\phi_A}\Delta V\ket{\phi_A}\Pbx=0$
(since $P_\xi^{\mathrm{int}}\PA\Delta v_{\xi A}\PA P_\xi^{\mathrm{int}}=0$
for each $\xi$);
it is retained in the general expression because a phenomenological
reference may not satisfy this condition
[cf.\ Eq.~(\ref{eq:correction}) in the main text].
The off-diagonal part
$\Pbx\bra{\phi_A}\Delta V\ket{\phi_A}\Qbx$, however, does not vanish
and enters $\Upol$ through the coupling
$\Pbx\Heff^{(A)}\Qbx$ in Eq.~(\ref{eq:Upol_full}) below.

Because the Feshbach elimination is performed sequentially (target first,
then projectile), the polarization term is constructed within the
target-projected effective Hamiltonian
$\Heff^{(A)}(E)=H_3^{(0)}+\Unonadd(E)$, which acts on the full $bx$-space
within $\PA$. Eliminating $\Qbx$ within this space gives
\begin{equation}\label{eq:Upol_full}
  \Upol(E) = \Pbx\Heff^{(A)}\Qbx\,
  \frac{1}{E{+}i\epsilon - \Qbx\Heff^{(A)}\Qbx}\,\Qbx\Heff^{(A)}\Pbx,
\end{equation}
which accounts for virtual excitation of projectile configurations
outside the $\Pbx$ model space and their subsequent de-excitation back into
$\Pspace$, with target-excitation effects already folded into the
propagator. Because $\Heff^{(A)}$ acts entirely within $\PA$-space, the
$\QA\Qbx$ sector that appears in the full resolvent
$\Qspace H\Qspace$ of Eq.~(\ref{eq:Heff}) is not double counted: it
is already integrated out through $\Unonadd$, and $\Upol$ only propagates
through $\PA\Qbx$ configurations. This term is the dynamical realization
of the spectroscopic factor discussed in the main text (where it is stated
that $\Upol$ ``encodes the physics traditionally compressed into a
spectroscopic factor''): its imaginary part describes flux loss into
channels not represented in $\Pspace$, while its real part shifts the
effective binding and scattering phase shifts. In the limit where $\Pbx$
retains only the projectile ground state, the norm of the overlap function
$\langle\phi_{bx}^{(0)}|\Psi_a\rangle$ and the dynamical effect of
$\Upol$ together account for the full projectile-structure content of
the reaction. For the $^{6}$Li benchmark, $d\alpha$ dominance implies
$\Upol\approx 0$ (the non-$d{+}\alpha$ configurations are dynamically
negligible), so the missing physics resides almost entirely in $\Unonadd$.

\textit{Exact channel-partition identity and controlled reduction.}---The
main text employs the two-term decomposition~(\ref{eq:decomposition}),
obtained by sequential elimination. Here I show that this form arises from
a more general three-term partition and is exact. Introducing three
orthogonal projectors in $\Qspace$:
\begin{equation}
  Q_1=\QA\Pbx,\qquad Q_2=\PA\Qbx,\qquad Q_3=\QA\Qbx,\qquad
  \Qspace=Q_1+Q_2+Q_3.
\end{equation}
Define $V_i=\Pspace H Q_i$, $W_i=Q_i H\Pspace$, and $H_{ij}=Q_i H Q_j$.
Then the induced interaction in Eq.~(\ref{eq:Heff}) is exactly
\begin{equation}\label{eq:Uind_block}
  U_{\mathrm{ind}}(E)=\sum_{i,j=1}^{3}
  V_i\bigl[(E{+}i\epsilon-H_{QQ})^{-1}\bigr]_{ij}W_j,
\end{equation}
with $H_{QQ}$ the $3\times 3$ block matrix $(H_{ij})$. Eliminating $Q_3$
by a Schur complement gives
\begin{equation}\label{eq:Uind_12}
  U_{\mathrm{ind}}(E)=\sum_{i,j=1}^{2}V_i\,\mathcal{G}_{ij}(E)\,W_j,
\end{equation}
where
\begin{equation}\label{eq:G12_def}
  \mathcal{G}^{-1}_{ij}(E)=\bigl(E{+}i\epsilon\bigr)\delta_{ij}-H_{ij}
  -H_{i3}\,G_3(E)\,H_{3j},
  \qquad
  G_3(E)=\frac{1}{E{+}i\epsilon-H_{33}}.
\end{equation}
Hence
\begin{align}\label{eq:exact_split}
  \Heff(E)=H_3^{(0)}
  +U_{\mathrm{nonadd}}^{\mathrm{ex}}(E)
  +U_{\mathrm{pol}}^{\mathrm{ex}}(E)
  +U_{\mathrm{mix}}^{\mathrm{ex}}(E),
\end{align}
with
\begin{align}
  U_{\mathrm{nonadd}}^{\mathrm{ex}}&=V_1\mathcal{G}_{11}W_1,\nonumber\\
  U_{\mathrm{pol}}^{\mathrm{ex}}&=V_2\mathcal{G}_{22}W_2,\nonumber\\
  U_{\mathrm{mix}}^{\mathrm{ex}}&=V_1\mathcal{G}_{12}W_2+V_2\mathcal{G}_{21}W_1.
\end{align}
The two-term decomposition~(\ref{eq:decomposition}) of the main text is
recovered via sequential elimination: first integrating out all of
$\QA$-space (including $Q_3$), which yields $\Unonadd$ with the full
$\QA$-resolvent, and then projecting onto $\Pbx$ using the
target-dressed Hamiltonian $\Heff^{(A)}$, which yields $\Upol$.
This sequential procedure is exact and avoids double counting of
$Q_3$; both the two-term~(\ref{eq:decomposition}) and
three-term~(\ref{eq:exact_split}) decompositions yield the same total
$\Heff$. The three-term form provides an alternative
bookkeeping; the mixing term $U_{\mathrm{mix}}^{\mathrm{ex}}$ measures the
coupling between target-excitation and projectile-truncation channels
that the sequential approach absorbs into $\Upol$ through the dressed
resolvent. A useful control parameter is
\begin{equation}\label{eq:eta_mix}
  \eta_{\mathrm{mix}}
  =\frac{\|U_{\mathrm{mix}}^{\mathrm{ex}}\|}
  {\|U_{\mathrm{nonadd}}^{\mathrm{ex}}\|+\|U_{\mathrm{pol}}^{\mathrm{ex}}\|},
\end{equation}
so $\eta_{\mathrm{mix}}\ll 1$ quantifies the regime where the two-term form
is numerically robust. For the $^{6}$Li system, $d\alpha$ dominance
implies weak coupling between the $Q_1$ ($\QA\Pbx$) and $Q_2$
($\PA\Qbx$) sectors, giving $\eta_{\mathrm{mix}}\ll 1$; the two-term
decomposition is therefore well justified in this case.

\textit{Operator-energy dependence and spectator shifts.}---When
the $\QA$-space resolvent is expanded in terms of the spectator
kinetic energy, the diagonal renormalization of the fragment--target
interactions acquires an explicit operator-energy structure:
\begin{equation}\label{eq:operator_energy}
  \mathcal{U}_{xA} \to \mathcal{U}_{xA}(E - K_b - \varepsilon_b),
  \quad
  \mathcal{U}_{bA} \to \mathcal{U}_{bA}(E - K_x - \varepsilon_x),
\end{equation}
where $K_b$ ($K_x$) is the kinetic-energy operator of the spectator
fragment and $\varepsilon_b$ ($\varepsilon_x$) its ground-state energy.
For the deuteron case ($a=d$, $b=n$, $x=p$), this structure was
conjectured by Austern and Richards~\cite{Austern1968} and formally
derived by Kozack and Levin~\cite{KozackLevin1986}; it underlies the
Johnson--Soper energy-shift prescription~\cite{Johnson1970} and its
generalizations to nonlocal
potentials~\cite{Timofeyuk2013PRC,Johnson2014,Bailey2017,Timofeyuk2020}.
For heavier projectiles the operator-energy dependence is qualitatively
similar but involves a broader spectator-momentum distribution, making
fixed-energy approximations less reliable.

\textit{Connection to the practical model Hamiltonian.}---Most
three-body reaction calculations employ the reduced model Hamiltonian
\begin{equation}\label{eq:HM}
  H_M = T_R + T_r + H_a + U_{bA} + U_{xA},
\end{equation}
which coincides with $H_3^{(0)}$ and therefore neglects both $\Unonadd$
and $\Upol$, as well as operator-energy corrections. The universal
framework derived here shows precisely what is dropped and provides a
path to systematic improvement: either by computing the induced terms
explicitly (where feasible), by fitting equivalent polarization
potentials to higher-fidelity calculations (e.g., four-body CDCC), or
by acknowledging the missing physics through model-dependent effective
spectroscopic factors. The $^{6}$Li benchmark demonstrates that using an
overcomplete reference potential (one that already absorbs part of the
induced interactions) without the compensating first-order
correction~(\ref{eq:correction}) double counts the absorption and
fails to reproduce elastic scattering data.
The three contributions to $R_s<1$ identified in the main text map
directly onto this framework: contribution~(i), genuine structure
depletion, corresponds to the overlap norm $S_P<1$
[Eq.~(\ref{eq:SP_def})]; contribution~(ii), overestimated $\sigma_{sp}$,
arises from the missing $\Unonadd$ and $\Upol$ in $H_M$
[Eq.~(\ref{eq:delta_sigma_neg})]; and contribution~(iii), model
inconsistency, corresponds to the nonvanishing first-order
correction~(\ref{eq:correction}) when a phenomenological potential that
already absorbs part of the induced physics is combined with dynamical
coupling that regenerates it.

\textit{Model-space spectroscopic strength.}---Let
$|\Psi^{(+)}\rangle$ be the exact scattering state of the full
Hamiltonian, $(E-H)|\Psi^{(+)}\rangle=0$, and define the projected component
$|\Psi^{P}\rangle=\Pspace|\Psi^{(+)}\rangle$. The Feshbach reduction yields
\begin{equation}\label{eq:Feshbach_equation}
  \bigl(E-\Heff(E)\bigr)|\Psi^{P}\rangle=0,
\end{equation}
so all flux leaving the model space is encoded in the imaginary parts of
$\Unonadd$ and $\Upol$. For a projectile ground state retained in $\Pbx$,
the spectroscopic strength for the $b{+}x$ partition is
\begin{equation}\label{eq:SP_def}
  S_{P}
  = \int d^3r\,|I_{bx}(\mathbf{r})|^2,
\end{equation}
with $I_{bx}$ the antisymmetrized overlap function of the projectile onto the
$b{+}x$ channel (integrated over fragment internal coordinates). A normalized overlap
$\tilde{I}_{bx}=I_{bx}/\sqrt{S_{P}}$ defines the channel wave function used
inside $\Pspace$.

Two consistent limits follow, corresponding to Route~A and Route~B of
the self-consistency criterion in the main text:

Route~A: if the reaction dynamics is computed with
$\Heff$ (or with an equivalent dynamical polarization potential), then the
loss of strength into excluded projectile configurations is already included
through $\Upol$, and the appropriate normalization is $S_{P}$; an additional
multiplicative SF would double count the same depletion. The factorized
cross section [Eq.~(\ref{eq:factorized})] becomes
$\sigma_{\mathrm{th}} = S_P\,\sigma_{sp}[\Heff]$, where
$\sigma_{sp}[\Heff]$ already reflects the absorption from the induced terms.

Route~B: if, instead, one
approximates the dynamics with $H_3^{(0)}$ by setting $\Upol\to 0$ (and
$\Unonadd\to 0$), then
the missing absorption inflates $\sigma_{sp}[H_3^{(0)}]$ relative to
$\sigma_{sp}[\Heff]$, and a phenomenological
overall reduction $S_{\mathrm{eff}} < S_P$ must be introduced to compensate.
The apparent quenching $R_s < 1$ in Route~B therefore reflects both genuine
depletion ($S_P < 1$) and dynamical incompleteness
($\sigma_{sp}[H_3^{(0)}] > \sigma_{sp}[\Heff]$). The extracted
$S_{\mathrm{eff}}$ is explicitly model dependent and should not be compared
directly to a structure-model $S_{P}$ computed in a different space.

\textit{Reference mismatch and first-order correction.}---Let
$U_{dA}^{F}$ denote a proper Feshbach reference satisfying the projected
condition
$P_d^{\mathrm{int}}\PA U_{dA}^{F}\PA P_d^{\mathrm{int}}
=P_d^{\mathrm{int}}\PA V_{dA}\PA P_d^{\mathrm{int}}$,
and let a phenomenological choice be
$U_{dA}^{\mathrm{ref}}=U_{dA}^{F}+\Delta U_{\mathrm{ref}}$. Then
\begin{equation}\label{eq:ref_mismatch}
  P_d^{\mathrm{int}}\PA\Delta v_{dA}\PA P_d^{\mathrm{int}}
  =P_d^{\mathrm{int}}\PA\!\left(V_{dA}-U_{dA}^{\mathrm{ref}}\right)\!\PA P_d^{\mathrm{int}}
  =-\Delta U_{\mathrm{ref}}.
\end{equation}
For $U_{dA}^{F}\approx U_d^{\mathrm{SF}}$ and
$U_{dA}^{\mathrm{ref}}=U_d^{\mathrm{OP}}$, this reduces to
Eq.~(\ref{eq:correction}) in the main text. The approximation
$U_{dA}^{F}\approx U_d^{\mathrm{SF}}$ identifies the bare Feshbach
reference with the single-folding potential; their difference, the
diagonal nucleon-target DPP, is absorbed into $U_d^{\mathrm{SF}}$
together with the corresponding diagonal second-order terms in
$\Unonadd$, leaving the $d$-breakup DPP as the dominant
uncompensated mismatch. The operator-energy correction to
$U_d^{\mathrm{SF}}$ from the spectator-momentum distribution is a
higher-order effect that has been quantified
in~\cite{Timofeyuk2013PRC,Bailey2017} and does not affect the
present argument.

\textit{Sufficient condition for $\delta\sigma_{sp}<0$.}---In the standard
eikonal stripping form,
\begin{equation}\label{eq:sigma_str}
  \sigma_{sp}^{\mathrm{str}}
  =\int d^2b\int d^3r\,\rho_{bx}(\mathbf{r})\,
  P_x^{\mathrm{rem}}(b,\mathbf{r})\,
  P_b^{\mathrm{surv}}(b,\mathbf{r}),
\end{equation}
with $P_x^{\mathrm{rem}}=1-|S_x|^2$ and $P_b^{\mathrm{surv}}=|S_b|^2$.
Here $S_b$ and $S_x$ are the eikonal $S$-matrices computed from the
additive reference potentials $U_{bA}$ and $U_{xA}$ in $H_3^{(0)}$
[Eq.~(\ref{eq:factorized})]; the induced terms $\Unonadd$ and $\Upol$
are omitted. Their imaginary parts provide additional absorption that
modifies the residue-survival probability. Modeling this net absorptive
contribution as an additional eikonal attenuation,
$P_b^{\mathrm{surv}}\to P_b^{\mathrm{surv}}e^{-\tau_{\mathrm{ind}}}$,
where $\tau_{\mathrm{ind}}(b,\mathbf{r})$ represents the
path-integrated imaginary part of the omitted interaction along the
residue trajectory. The modification of $P_x^{\mathrm{rem}}$ is
neglected at this order; for deeply bound removal this is justified
because $P_x^{\mathrm{rem}}\approx 1$ in the interaction region,
leaving little room for further increase. When
$\tau_{\mathrm{ind}}\ge 0$,
\begin{equation}\label{eq:delta_sigma_neg}
  \delta\sigma_{sp}^{\mathrm{str}}
  =-\!\int d^2b\!\int d^3r\,\rho_{bx}\,
  P_x^{\mathrm{rem}}P_b^{\mathrm{surv}}
  \left(1-e^{-\tau_{\mathrm{ind}}}\right)\le 0.
\end{equation}
For weak induced attenuation,
\begin{equation}\label{eq:delta_sigma_lin}
  \delta\sigma_{sp}^{\mathrm{str}}
  =-\!\int d^2b\!\int d^3r\,\rho_{bx}\,
  P_x^{\mathrm{rem}}P_b^{\mathrm{surv}}\,
  \tau_{\mathrm{ind}}+O(\tau_{\mathrm{ind}}^2).
\end{equation}
Thus the additive model overestimates $\sigma_{sp}$ whenever the net
induced absorption in the residue-survival channel is positive.
The optical theorem applied to the model-space $S$~matrix requires
net flux loss into excluded channels whenever the collision energy
exceeds the $\Qspace$-space thresholds, as is standard for
intermediate-energy knockout. While the local imaginary part of the
induced interaction need not be negative everywhere, its
spatially averaged effect along the dominant trajectories is
absorptive, so $\langle\tau_{\mathrm{ind}}\rangle > 0$ and
$\delta\sigma_{sp}^{\mathrm{str}} < 0$ in the net. This result underpins the
additive-model side of the argument in the main text: the standard
knockout analysis uses $H_3^{(0)}$ without the induced interactions,
leading to $\delta\sigma_{sp}<0$ and consequently $R_s<1$.

\textit{Relative importance of $\Unonadd$ and $\Upol$ for nucleon
knockout.}---The $^{6}$Li benchmark is governed by $\Unonadd$: the
$d{+}\alpha$ model space captures ${\sim}70\%$ of the $^{6}$Li wave
function and, more importantly, the non-$d{+}\alpha$ configurations
($\Qbx$-space) couple negligibly to the elastic channel ($d\alpha$
dominance~\cite{Watanabe2015}), so $\Upol\approx 0$. For standard
nucleon knockout ($a=\mathrm{core}+N$), the situation is qualitatively
different. The model space $\Pbx$ retains only the core--nucleon
ground-state configuration; $\Qbx$ encompasses all core-excited states,
the full $\mathrm{core}{+}N$ continuum beyond the retained bins, and all
non-cluster rearrangement channels. The spectroscopic factor
$S_P\approx 0.5$--$0.7$ leaves $30$--$50\%$ of the projectile wave
function in $\Qbx$, and, unlike the $^{6}$Li case, core excitations
are known to couple strongly to the reaction dynamics (core-excitation
DPPs are substantial~\cite{Moro2012}). Consequently $\Upol$ is expected
to be the leading missing interaction for standard knockout, with
$\Unonadd$ providing a secondary correction.

This inversion has important practical consequences.
First, since $\Upol$ [Eq.~(\ref{eq:Upol_full})]
is constructed from the \textit{target-dressed} Hamiltonian
$\Heff^{(A)}$, it is an irreducibly three-body operator: it
depends simultaneously on the nucleon--target and core--target
coordinates and cannot be absorbed into a correction to either
$U_{NA}$ or $U_{\mathrm{core},A}$ alone. In the eikonal framework,
this means the induced term breaks the standard factorization
$S_{\mathrm{tot}}(\mathbf{b}_N,\mathbf{b}_{\mathrm{core}})
=S_N(\mathbf{b}_N)\,S_{\mathrm{core}}(\mathbf{b}_{\mathrm{core}})$;
a faithful representation requires a correlated $S$-matrix
$S_{\mathrm{tot}}=S_N S_{\mathrm{core}}\,
S_{\mathrm{ind}}(\mathbf{b}_N,\mathbf{b}_{\mathrm{core}},\mathbf{r})$
that encodes the geometry-dependent induced interaction.
Second, the imaginary part of $\Upol$ is guaranteed to be absorptive
above the $\Qbx$ thresholds by the optical theorem (it represents net
flux loss into excluded projectile configurations); however, its real
part can be either attractive or repulsive, and its net effect on
$\sigma_{sp}$ includes not only the survival-probability reduction
discussed in Eq.~(\ref{eq:delta_sigma_neg}) but also modifications of
the scattering wave function itself. A reliable estimate of the total
correction therefore requires a dynamical calculation, whether through
explicit channel coupling or through calibrated equivalent
polarization potentials, rather than a simple absorptive
parameterization.

These considerations underscore why a direct quantitative evaluation of
the induced terms for nucleon knockout remains an open problem:
$\Unonadd$ requires a microscopic treatment of correlated
nucleon--target and core--target virtual excitations, while $\Upol$
demands coupling to the full $\Qbx$-space of the projectile within a
target-dressed propagator. Both are many-body problems intractable with
present methods. The $^{6}$Li benchmark, the only system where all
induced interactions can be evaluated through a controlled four-body
calculation, validates the formalism and the mechanism; extending the
quantitative assessment to heavier projectiles is the central challenge
for future work.

\textit{Mapping induced terms to equivalent polarization potentials.}---For
practical calculations it is useful to replace the nonlocal operators
$\Unonadd$ and $\Upol$ by equivalent polarization potentials acting in a
reduced channel space. Let $P_0=\PA\ket{\phi_{bx}^{(0)}}\bra{\phi_{bx}^{(0)}}$
project onto the elastic channel within $\Pspace$, and
$Q_0=\Pspace-P_0$ the remaining channels (e.g., projectile continuum bins in
CDCC). A second Feshbach reduction within $\Pspace$ yields the exact elastic
equation
\begin{equation}\label{eq:U_DPP}
  \bigl[T_R+U_{bA}+U_{xA}+U_{\mathrm{DPP}}(E)-E\bigr]\chi_0(\mathbf{R})=0,
\end{equation}
with the dynamical polarization potential
\begin{equation}\label{eq:U_DPP_def}
  U_{\mathrm{DPP}}(E)=P_0\,(\Unonadd+\Upol)\,P_0
  + P_0 H_{\mathrm{eff}}\,Q_0\frac{1}{E{+}i\epsilon-Q_0 H_{\mathrm{eff}}Q_0}
  Q_0 H_{\mathrm{eff}}\,P_0 .
\end{equation}
This operator is nonlocal and energy dependent; a local or separable
representation can be obtained by requiring that it reproduce the elastic
$S$ matrix (or phase shifts) from a higher-fidelity calculation. In practice,
one can define an equivalent local potential $U_{\mathrm{DPP}}^{\mathrm{loc}}$
by $S$-matrix inversion of four-body CDCC results and then use
$U_{bA}+U_{xA}+U_{\mathrm{DPP}}^{\mathrm{loc}}$ in a reduced three-body
model. The difference between three-body calculations that use
$U_d^{\mathrm{OP}}$ and those that use $U_d^{\mathrm{SF}}$ is a direct
manifestation of the first-order non-additive
correction~(\ref{eq:correction}), and benchmark
four-body\,$\rightarrow$\,three-body mappings provide a controlled route to
quantify both this term and the higher-order contributions to
$U_{\mathrm{DPP}}$.

\end{document}